\documentclass[aps,apl,twocolumn,superscriptaddress]{revtex4}
\usepackage{graphicx}
\usepackage{epstopdf}
\usepackage{comment}
\begin{document}

\title{Mode-sensitive magnetoelastic coupling in phononic-crystal magnomechanics}

\author{D. Hatanaka}

\email{daiki.hatanaka.hz@hco.ntt.co.jp}


\author{H. Yamaguchi}

\affiliation{NTT Basic Research Laboratories, NTT Corporation, Atsugi-shi, Kanagawa 243-0198, Japan}

\begin{abstract}
Acoustically driven spin-wave resonance in a phononic crystal cavity is numerically investigated. The designed cavity enables confinement of gigahertz vibrations in a wavelength-scale point-defect structure and sustains a variety of resonance modes. Inhomogeneous strain distributions in the modes modify the magnetostrictive coupling and the spin-wave excitation susceptible to an external field orientation. In particular, a monopole-$\it{like}$ mode in the cavity having a near-symmetrical pattern shows a subwavelength-scale mode volume and can provide a versatile acoustic excitation scheme independent on field-angle variation. Thus, the phononic-crystal platform offers an alternative approach to acoustically control the spin-wave dynamics with ultrasmall and inhomogeneous mode structures, which will be a key technology to integrate and operate large-scale magnomechanical circuits.
\end{abstract} 

\maketitle

 \begin{figure}[t]
	\begin{center}
		\vspace{-0.0cm}\hspace{0.8cm}
		\includegraphics[scale=0.86]{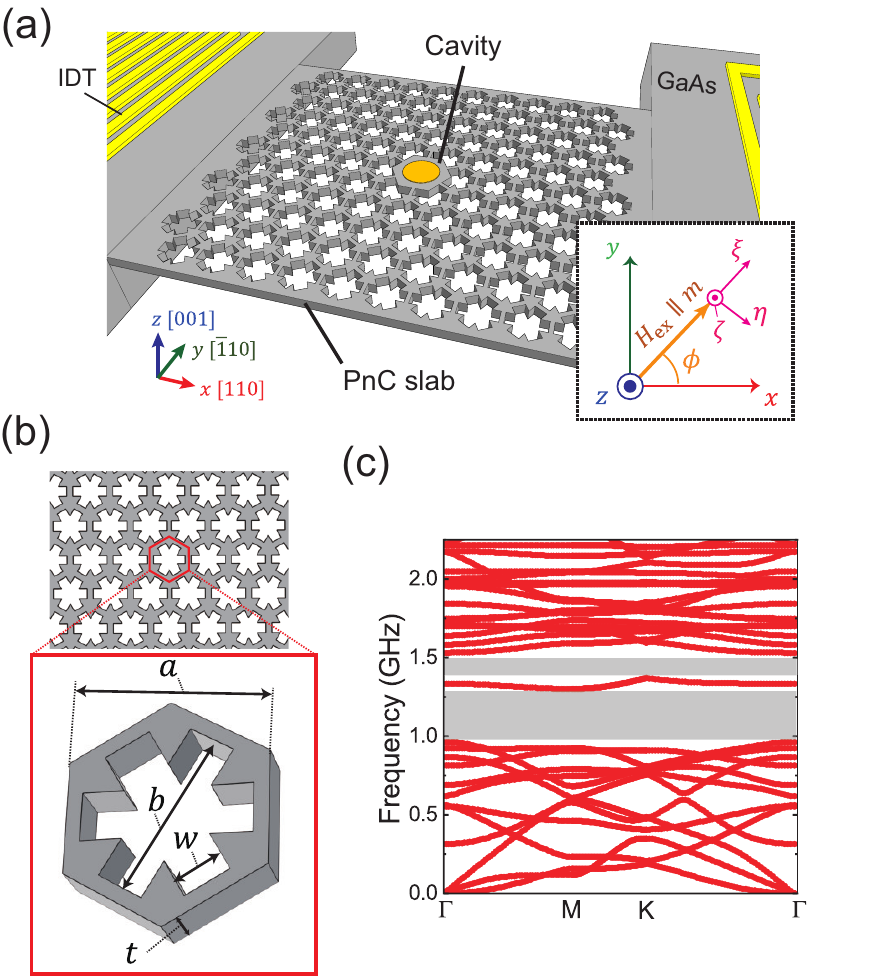}
		\vspace{-0cm}
		\caption{\textbf{(a)} Illustration of PnC-based magnomechanics in which a ferromagnet thin film (Ni) with thickness $d_{\rm Ni}$ = 50 nm is formed on the surface of a point-defect cavity in a suspended membrane. This phononic cavity can be driven by an electromechanical transducer such as an IDT (inter-digit transducer). The crystal orientations of GaAs are denoted as [110], [$\bar{1}$10] and [001]. GaAs is chosen as the PnC medium because the microfabrication technology for it is well developed and it hosts the piezoelectricity enabling on-chip driving of coherent vibrations through the IDT. The inset shows the relationship between the $xyz$- and $\zeta\eta\xi$-coordinate systems. \textbf{(b)} Schematic of the PnC geometry consisting of triangular lattice of a snowflake-shaped air-hole. The inset shows the unit structure having dimensions $a$ = 2.0 $\mu$m, $b$ = 1.7 $\mu$m, $w$ = 0.5 $\mu$m and $t$ = 0.5 $\mu$m. \textbf{(c)} Dispersion relation of acoustic waves in the PnC, showing full bandgaps in the frequency range as highlighted in gray. The bandgaps are divided by a phonon branch sustaining edge vibrations of the unit structure. All the resonance modes studied in this work exist in the lower bandgap and an alternative mode is also found in the higher one. Further discussions about the dispersion relation are given in supplementary material.}
		\label{fig 1}
		\vspace{-0.7cm}
	\end{center}
\end{figure}
\hspace*{0.5em}Mutual coupling between acoustic waves and magnetic moments in a ferromagnet has been intensively investigated over the last several years \cite{weiler_adfmr,dreher_adfmr,weiler_spinpump,matsuo_src,thevenard_adfmr1,labanowski_apl,labanowski_adfmr,kobayashi_src,thevernard_prap2018,onose_PRB2019,santos_magnomech2020,otani_nnano2020,otani_iop2020,otani_sciad2020,yu_prl2020,kikkawa_prl2016,saitoh_ncom2017,schmidt_ncom2019,saitoh_ncom2019,klein_prb2020,whiteley_nvspin}. Developing the acoustic means to drive ferromagnetic and spin-wave resonances is increasingly important to explore the possibility of phononics technology in the field of spintronics. Previous magnomechanical systems are mostly constructed on a surface acoustic-wave (SAW) device, in which excited acoustic plane waves permit the dynamics of magnon-phonon interaction to be studied with a simple analytical model \cite{weiler_adfmr,dreher_adfmr,weiler_spinpump,matsuo_src,thevenard_adfmr1,labanowski_apl,labanowski_adfmr,kobayashi_src,thevernard_prap2018,onose_PRB2019,santos_magnomech2020,otani_nnano2020,otani_iop2020,otani_sciad2020,yu_prl2020}. However, SAW-based driving via the magnetostriction \cite{weiler_adfmr,dreher_adfmr} and Barnett effect \cite{matsuo_src,kobayashi_src} only work in restricted external-field orientations. In addition, the SAW energy loss to a bulky substrate results in poor spatial confinement \cite{loncar_sawpnc,tang_sawpnc}, and the subsequent large vibrational mode volume reduces the spatial power density and weakens the effective interaction. Thus, an alternative and versatile platform for magnomechanics is highly desired.\\
\hspace*{0.5cm}In this letter, we propose to use a phononic crystal (PnC) acoustic cavity for controlling the magnomechanical interaction. The study is based on a numerical investigation of the resonant modal structures and the characteristics of acoustically-driven spin-wave resonance at various field angles. The phononic bandgap formed by the periodic structure fully confines gigahertz vibrations in the wavelength-scale point-defect structure. This structure resonates at specific frequencies, revealing a variety of modal shapes. The resultant strains are inhomogeneously distributed so that the field response of power absorption due to the spin-wave excitation is variant with respect to the resonant mode structures. Dipole- and quadrupole-$\it{like}$ modes clearly show different field-angle dependency, whereas the field-angle variation mostly vanishes in a monopole-$\it{like}$ mode. The PnC-based magnomechanics is promising for spatially manipulating hypersonic waves and tailoring the spin-wave excitation efficiency with respect to the field direction and thus for enhancing the directionality and functionality of magnomechanical elements.\\
\hspace*{0.5cm}The simulated PnC is constructed by a triangular lattice of snowflake-shaped air holes formed in a suspended GaAs slab as shown in Fig. 1(a) and the original design has been proposed by Safavi-Naeini $\textit{et al}$ \cite{safavi_snowflake0}. By designing the periodic structure in Fig. 1(b), a complete bandgap is formed in the frequency ranges of 0.97-1.30 GHz and 1.37-1.53 GHz, and the dispersion relation is shown in Fig. 1(c). A point-defect structure is created by removing one showflake hole from the lattice, and it resonates at frequencies within the bandgap. This geometry is chosen as a PnC cavity because we have confirmed the fabrication and experimental possibilities previously and the details on the mechanical properties are found in elsewhere \cite{hatanaka_hyperPnC}. A circular polycrystalline nickel (Ni) film is formed on the defect. Finite-element method (FEM) calculation using COMSOL Multiphysics reveals that multiple Lamb- and Love-type resonance modes are formed in the cavity. Here, we focus on three asymmetric Lamb modes, labeled monopole-, dipole- and quadrupole-$\it{like}$ modes, and investigate the variation in the magnetostriction with changing external field orientation.\\
\hspace*{0.5cm}Acoustic spin-wave excitation by the PnC cavity is simulated through the strain distribution of the resonant modal shapes. From various strain components $\epsilon_{\rm ij}$ ($i,j = x,y,z$), effective fields ($h_{\rm \zeta}$, $h_{\rm \eta}$) via magnetostriction are given by \cite{dreher_adfmr},
\begin{eqnarray}
\mu_{0} h_{\rm \zeta} &=& 2 b_{\rm s} \left( \epsilon_{\rm xz} \cos{\phi} + \epsilon_{\rm yz} \sin{\phi} \right), \\
\mu_{0} h_{\rm \eta} &=& 2 b_{\rm l} \sin{\phi} \cos{\phi} \left( \epsilon_{\rm xx} - \epsilon_{\rm yy} \right) -2 b_{\rm s} \epsilon_{\rm xy} \cos{2 \phi},
\end{eqnarray}
with the shear and longitudinal magnetostrictive coupling constants $b_{\rm s}$ and $b_{\rm l}$ respectively, and vacuum permeability $\mu_{0}$. An alternative $\zeta\eta\xi$-coordinate system is defined in such a way that the $\xi$-axis is aligned to the magnetization whose angle is set to $\phi$ with respect to $x$-axis as shown in the inset of Fig. 1(a). Applying time-varying strains to the cavity via an IDT generates magnetization precession acoustically without rf electromagnetic waves, which is the usual driving technique for ferromagnetic resonance experiments. This also results in acoustic power being absorbed, which suppresses and modulates the resonance amplitude and phase. Using (1) and (2), the change in the power via the magnetostrictive driving ($\Delta P$) is expressed by
\begin{equation}
\Delta P = -\frac{\omega \mu_{0}}{2} \int_{V_{0}} \left[ \left(h_{\rm \zeta}^{*}, h_{\rm \eta}^{*} \right) \bar{\chi}
\left(
\begin{array}{c}
h_{\rm \zeta} \\
h_{\rm \eta} \\
\end{array}
\right)
\right] dV
\end{equation}
where $\omega$ is the angular frequency of acoustic waves and $V_{0}$ is the volume of Ni film. The magnetic susceptibility $\bar{\chi}$ is obtained from the magnetic free energy $G = -\mu_{0} $\mbox{\boldmath $H_{\rm ex}$}$ \cdot  $\mbox{\boldmath $m$}$ + B_{\rm d} m_{\rm z}^{2}$ with the external static field $\mu_{0} H_{\rm ex}$, out-of-plane shape anisotropy for the thin film $B_{\rm d}=\mu_{0}M_{\rm s}/2$, unit magnetization $\mbox{\boldmath $m$}=\mbox{\boldmath $M$}/M_{\rm s}$, and saturation magnetization $M_{\rm s}$. Here, we assume that the static field is applied in the plane of the Ni film and ignore the in-plane anisotropy for simplicity due to the polycrystalline and circular structures. Then, the available components of $\bar{\chi}$ are given by $\chi_{11} = \gamma \mu_{0}M_{\rm s} \left( \gamma \mu_{0}H_{\rm ex} - i\omega\alpha \right)/D$, $\chi_{12} = \chi_{21}^{*} = -i\gamma\mu_{0}M_{\rm s}\omega/D$ and $\chi_{22} = \gamma\mu_{0}M_{\rm s} \left[ \gamma \left( 2B_{\rm d} + \mu_{0} H_{\rm ex} \right) - i\omega\alpha \right]$ with $D = \left[ \gamma \left( 2B_{\rm d} + \mu_{0}H_{\rm ex} \right) -i\omega\alpha \right] \left( \gamma\mu_{0}H_{\rm ex} - i\omega\alpha \right) - \omega^{2}$, where $\gamma$ is the gyromagnetic ratio and $\alpha$ is the Gilbert damping ratio. The detailed derivation and explanation of the above formula are shown elsewhere \cite{dreher_adfmr}. Thus, taking the real and imaginary parts of $\Delta P$ gives variation in the dispersion ($P_{\rm d}$) and attenuation ($P_{\rm a}$) by the spin-wave excitation. In the simulations, the acoustic parameters for GaAs were density $\rho$ = 5360 kg/m$^{3}$, elastic constants $C_{11}$ = 111.8 GPa, $C_{12}$ = 53.8 GPa, $C_{44}$ = 59.4 GPa, and mechanical (acoustic) loss-factor $Q^{-1}$ = $5 \times 10^{-4}$. These acoustic parameters are experimentally valid from previous works so that defect and randomness in the structure and material property during device fabrication are considered in the calculation \cite{iwamoto_topo2019,hatanaka_hyperPnC}. For Ni, they were density $\rho$ = 8900 kg/m$^{3}$, Young modulus $E$ = 219 GPa and Poisson's ratio $\nu$ = 0.31. The magnetic parameters from previous work were used: $b_{\rm s}=b_{\rm l}$ = 23 T (polycrystalline), $B_{\rm d}$ = 0.2 T, $M_{\rm s}$ = 370 kA/m, and $\gamma$ = 2.185$\mu_{\rm B}/\hbar$ with Bohr's magneton $\mu_{\rm B}$ and the reduced Plank constant $\hbar$ and $\alpha$ = 0.05.\\
\begin{figure}[t]
	\begin{center}
		\vspace{-0.2cm}\hspace{-0.0cm}
		\includegraphics[scale=0.9]{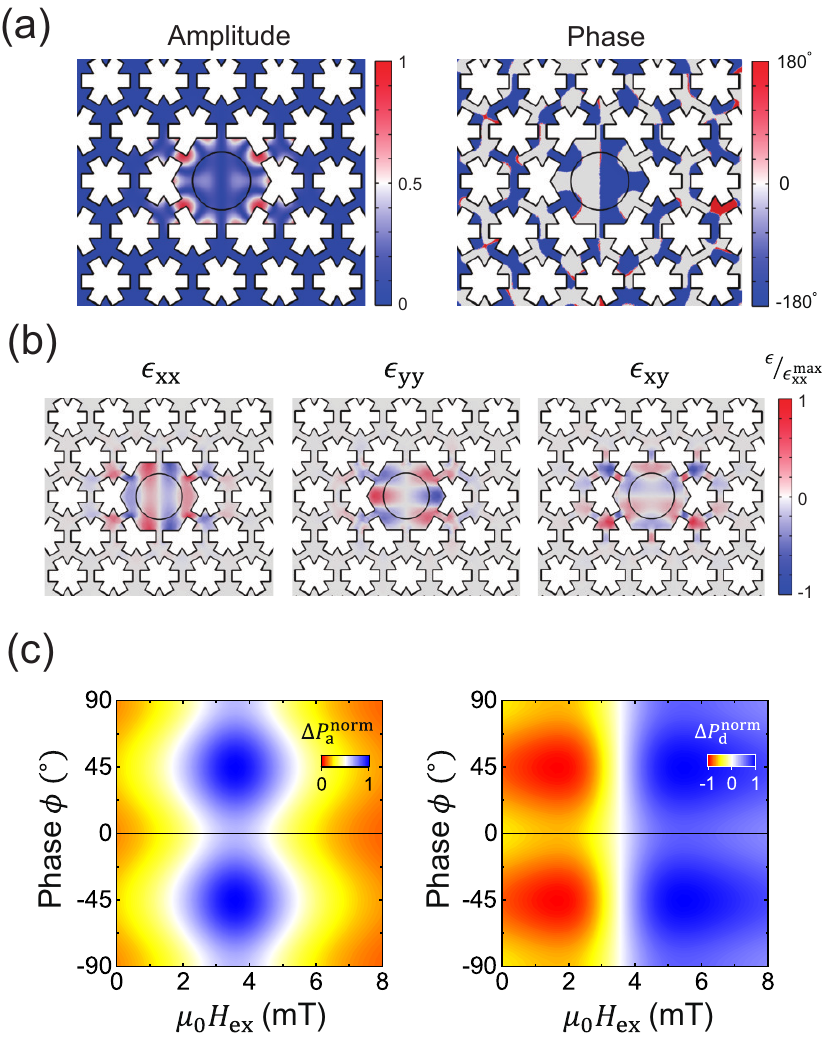}
		\vspace{-0cm}
		\caption{\textbf{(a)} Vibration amplitude and phase of the dipole mode at $\omega/2\pi$ = 1.160 GHz. \textbf{(b)} From left to right, spatial distribution of the strains $\epsilon_{\rm xx}$, $\epsilon_{\rm yy}$ and $\epsilon_{\rm xy}$, which are normalized by the maximum strain of $\epsilon_{\rm xx}^{\rm max}$. \textbf{(c)} Field dependence of the attenuation (left) and dispersion (right) of the acoustic resonant vibrations at various $\phi$ between -90$^\circ$ and 90$^\circ$ absorbed by spin-wave excitation via the magnetostriction. The maximum attenuation is $|P_{\rm a}|_{\rm max}$ = 5.0 nW in $\epsilon_{\rm xx}^{\rm max} = 4.3 \times 10^{-5}$.}
		\label{fig 2}
		\vspace{-0.7cm}
	\end{center}
\end{figure}
\begin{figure}[t]
	\begin{center}
		\vspace{-0.2cm}\hspace{-0.0cm}
		\includegraphics[scale=0.9]{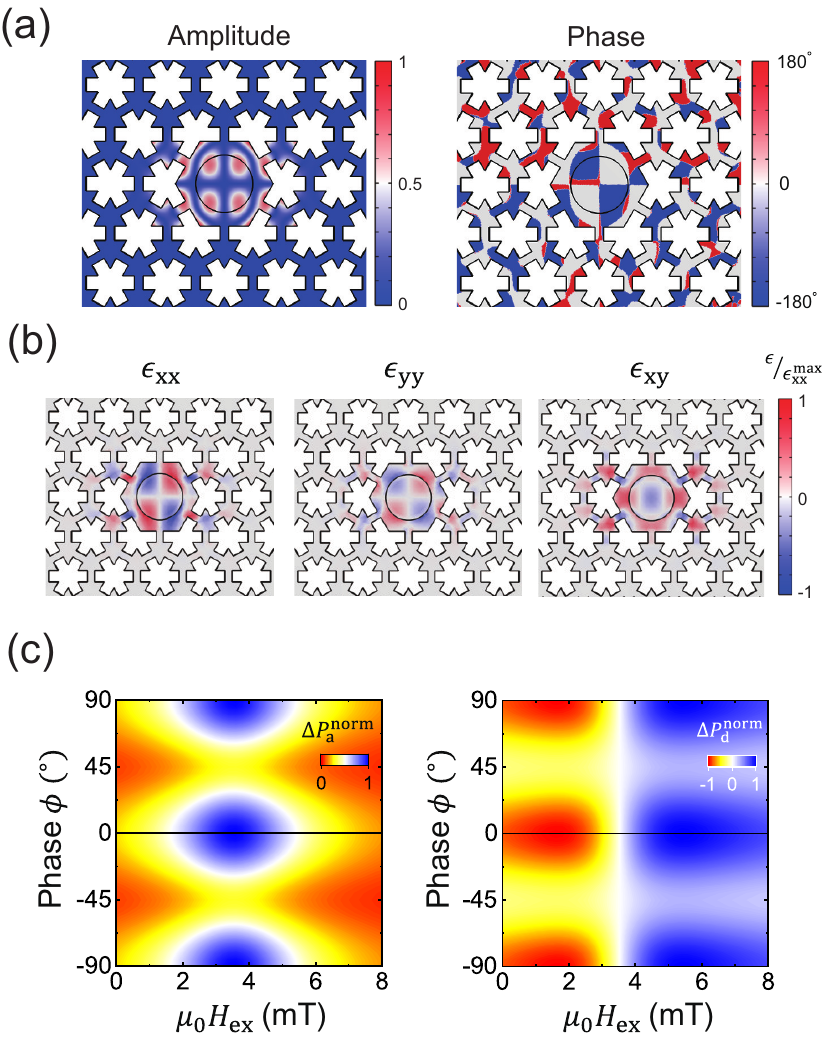}
		\vspace{-0cm}
		\caption{\textbf{(a)} Vibration amplitude and phase of the quadrupole mode structure at $\omega/2\pi$ = 1.156 GHz. \textbf{(b)} From left to right, spatial distribution of the strains $\epsilon_{\rm xx}$, $\epsilon_{\rm yy}$ and $\epsilon_{\rm xy}$, which are normalized by the maximum strain of $\epsilon_{\rm xx}^{\rm max}$. \textbf{(c)} Field dependence of the attenuation (left) and dispersion (right) of the acoustic resonant vibrations at various $\phi$ between -90$^\circ$ and 90$^\circ$ absorbed by spin-wave excitation via the magnetostriction. The maximum attenuation is $|P_{\rm a}|_{\rm max}$ = 7.7 nW in $\epsilon_{\rm xx}^{\rm max} = 4.4 \times 10^{-5}$.}
		\label{fig 3}
		\vspace{-0.7cm}
	\end{center}
\end{figure}
\begin{figure}[t]
	\begin{center}
		\vspace{-0.2cm}\hspace{-0.0cm}
		\includegraphics[scale=0.9]{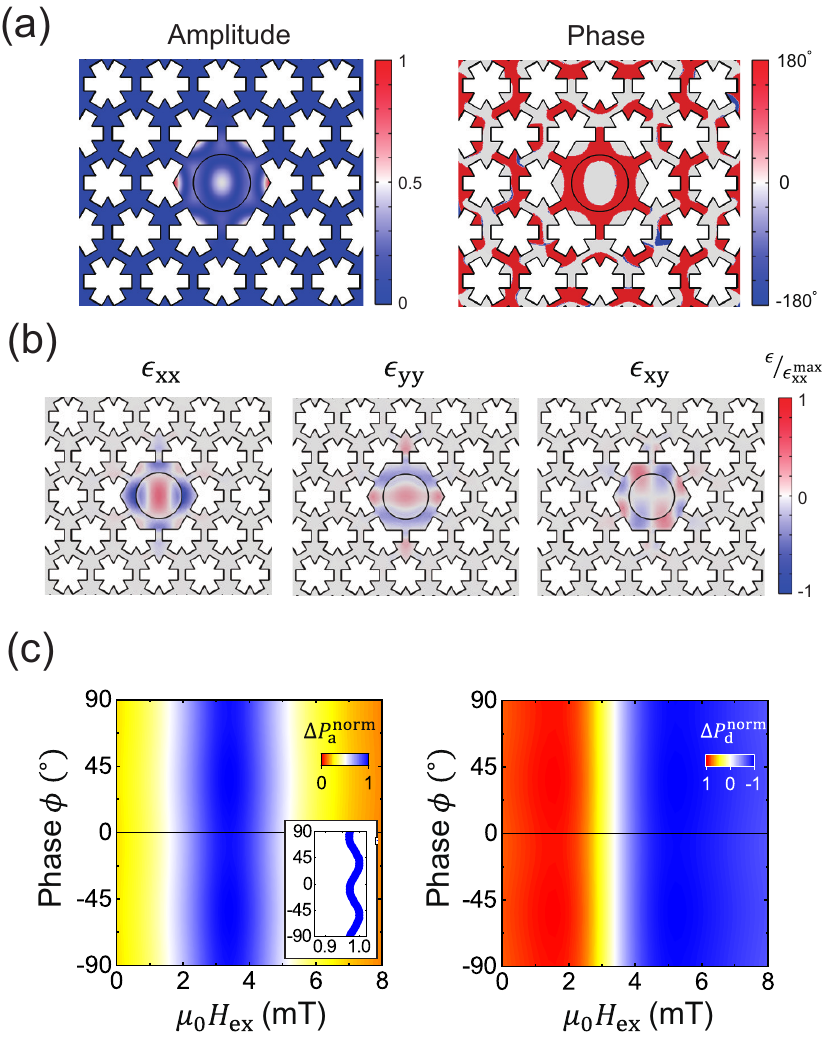}
		\vspace{-0cm}
		\caption{\textbf{(a)} Vibration amplitude and phase of the monopole mode structure at $\omega/2\pi$ = 1.131 GHz as shown in the left and right panels respectively. \textbf{(b)} From left to right, spatial distribution of the strains $\epsilon_{\rm xx}$, $\epsilon_{\rm yy}$ and $\epsilon_{\rm xy}$, which are normalized by the maximum strain of $\epsilon_{\rm xx}^{\rm max}$. \textbf{(c)} Field dependence of the attenuation (left) and dispersion (right) of the acoustic resonant vibrations at various $\phi$ between -90$^\circ$ and 90$^\circ$ absorbed by spin-wave excitation via the magnetostriction. The maximum attenuation is $|P_{\rm a}|_{\rm max}$ = 1.9 nW in $\epsilon_{\rm xx}^{\rm max} = 3.3 \times 10^{-5}$. The inset in the left panel indicates the angle dependence of $P_{\rm a}^{\rm norm}$ at $\mu_{0} H_{\rm ex}$ = 3.4 mT.}
		\label{fig 4}
		\vspace{-0.7cm}
	\end{center}
\end{figure}
\hspace*{0.0cm}First, a dipole-$\it{like}$ cavity mode is considered. The vibration amplitude and phase of the modal shape are displayed in the left and right panels of Fig. 2(a), respectively, showing that two anti-nodes oscillate out of phase at $\omega$/(2$\pi$) = 1.160 GHz. This enables the effective mode volume to be calculated by integrating vibration energy over the slab as $V_{\rm eff} = \int d\mbox{\boldmath $r$}^{2} \left( \frac{|u(\mbox{\boldmath $r$})|}{|u(\mbox{\boldmath $r$})|_{\rm max}}\right)^{2}t$ = 0.852 $\mu$m$^{3}$ $\approx$ 0.57$\lambda^{2}t$, where $u(\mbox{\boldmath $r$})$ ($|u(\mbox{\boldmath $r$})|_{\rm max}$) is vibration (maximum) displacement at position $\mbox{\boldmath $r$}$, and $\lambda$ is the acoustic wavelength \cite{hatanaka_hyperPnC}. The result indicates the strong confinement of gigahertz vibrations in the tiny sub-wavelength-scale space. The surface strain distributions were also calculated, and only the strain components available for the magnetostrictive driving, $\epsilon_{\rm xx}$, $\epsilon_{\rm yy}$ and $\epsilon_{\rm xy}$ are shown in Fig. 2(b), whose magnitudes are normalized by the maximum $\epsilon_{\rm xx}^{\rm max}$. In contrast to usual SAW-based magnomechanical systems where only longitudinal strain such as $\epsilon_{\rm xx}$ is assumed to be nonzero, this PnC cavity has a two-dimensionally confined structure so that additional longitudinal and shear strains such as $\epsilon_{\rm yy}$ and $\epsilon_{\rm xy}$ are involved. Out-of-plane shear strains $\epsilon_{\rm xz}$ and $\epsilon_{\rm yz}$ vanish due to the surface boundary condition \cite{morgan_saw}. As a result, $\mu_{0}h_{\rm \zeta}$ = 0 and only $\mu_{0}h_{\rm \eta}$ contribute to the spin-wave excitation. The calculated strains are substituted into (2) and then, the acoustic power absorption is obtained through (3). The resultant dispersion and attenuation, normalized by the maximum values $P_{\rm d}^{\rm norm} = P_{\rm d}(H_{\rm ex},\phi)/|P_{\rm d}(H_{\rm ex},\phi)|_{\rm max}$ and $P_{\rm a}^{\rm norm} = P_{\rm a}(H_{\rm ex},\phi)/|P_{\rm a}(H_{\rm ex},\phi)|_{\rm max}$, are plotted as a function of the field at $\phi$ ranging from -90$^{\circ}$ to 90$^{\circ}$ as shown in the left and right panels of Fig. 2(c) respectively. As in conventional ferromagnet-SAW systems \cite{weiler_adfmr,dreher_adfmr,labanowski_apl,labanowski_adfmr}, the largest attenuation and dispersion shift occur at $\phi  = \pm 45^\circ$. This PnC still yields small but observable spin-wave absorptions around $\phi$ = 0$^\circ$ and $\pm$90$^\circ$, at which no absorption occurs in SAW-based systems. This behavior can be understood from (2), where the shear strain $\epsilon_{\rm xy}$ varies with $\cos{2\phi}$ and thus the magnetostriction is available even at these angles. This investigation reveals that the PnC cavity with the inhomogeneous strains is capable of driving spin precession acoustically.\\
\hspace*{0.5cm}To investigate the variation in the field-angle response when changing the modal shape, the spin-wave-induced absorption in the quadrupole-$\it{like}$ mode with $\omega/2\pi$ = 1.156 GHz was calculated in attenuation and dispersion. This mode has four vibration anti-nodes, each of which oscillates out of phase with respect to neighboring ones as shown in Fig. 3(a). The effective mode volume is calculated as $V_{\rm eff}$ = 1.007 $\mu$m$^{3}$ $\approx$ 0.68$\lambda^{2}t$. The resultant strains $\epsilon_{\rm xx}$, $\epsilon_{\rm yy}$ and $\epsilon_{\rm xy}$ shown in Fig. 3(b) induce the normalized acoustic attenuation and dispersion via the magnetostriction. The field dependencies are shown in the left and right panels of Fig. 3(c), respectively. Remarkably, the susceptibility to the field angle changes from that of the dipole mode, resulting in the maximum absorption at $\phi$ = 0$^\circ$ and $\pm$90$^\circ$, whereas the absorption is small at $\phi$ = $\pm$45$^\circ$. The result can be interpreted to mean that the shear strain $\epsilon_{\rm xy}$ dominates the magnetostrictive field from (2), whereas the difference in the longitudinal strains, $\epsilon_{\rm xx}-\epsilon_{\rm yy}$ in (2), is reduced. Thus, different modal structures enable us to modify the magnetostrictive characteristics of the PnC.\\
\hspace*{0.5cm}The point-defect structure in the cavity sustains the monopole-$\it{like}$ mode having a hexagonal modal shape with $\omega/2\pi$ = 1.131 GHz and the smallest $V_{\rm eff}$ = 0.281 $\mu$m$^{3}$ $\approx$ 0.19$\lambda^{2}t$, as shown in Fig. 4(a). This hosts nearly radial symmetry on the Ni thin film and thus the generated strains shown in Fig. 4(b), although six-fold rotational symmetry slightly remains in the vibrational structure close to the defect edges. Therefore, the simulation reveals that the acoustic modulations, namely the acoustic driving efficiency of the spin-wave resonance, are mostly insensitive to variation in the field angle as shown in Fig. 4(c). Indeed, there is a little fluctuation in the maximum of $P_{\rm a}^{\rm norm}$ while $\phi$ is swept as shown in the inset of Fig. 4(c), but it is negligibly small compared to that of other modes and SAW-based magnomechanical systems. This monopole mode inherent to the point-defect cavity is useful for developing and operating PnC-based magnomechanical circuitry because it offers compact and versatile control of the spin precession regardless of the external field angles.\\
\hspace*{0.5cm}In conclusion, we numerically investigated the magnomechanical characteristics of a point-defect cavity in a suspended PnC slab. A variety of resonant modal shapes emerge in the wavelength-scale cavity sustained by the bandgap, resulting in the formation of inhomogeneous strain distributions. The spatial variation in the longitudinal and shear strains in the modes determines the field-angle dependence of the magnetostriction, namely the excitation efficiency of spin-wave resonance. The opposite dependency with respect to the orientation is found between the dipole and quadruple modes. It is remarkable that the monopole mode is mostly insensitive to the field orientation for the spin-wave excitation. Thus, the PnC architecture is a promising platform for enhancing the directionality and versatility of the magnetostrictive driving scheme, which holds promise for developing hybrid magnomechanical circuitry.\\

\vspace*{0.1cm}
\hspace*{-0.35cm}\textbf{SUPPLEMENTARY MATERIAL}\\
\hspace*{0.5cm}See supplementary material for the dispersion relation of the phononic branch between the badngaps and the magnetostrictive effect on the edge resonant mode.

\vspace*{0.5cm}
\hspace*{-0.35cm}\textbf{ACKNOWLEDGEMENTS}\\
\hspace*{0.5cm}The authors thank H. Okamoto, Y. Kunihashi and H. Sanada for fruitful discussion.

\vspace*{0.5cm}
\hspace*{-0.35cm}\textbf{DATA AVAILABILITY}\\
\hspace*{0.5cm}The data that support the findings of this study are available from the corresponding author upon reasonable request.


\end{document}